# Re-refinement of the structure of the planar hexagonal phase of ZnO nanocrystals

**Musen Li.[abc] Lingyao Zhang.[ad] Wei Ren.[ab*] and Jeffrey R. Reimers[ac*]**

[a] International Centre for Quantum and Molecular Structures and Department of Physics, Shanghai University, Shanghai 200444, China.

[b] Materials Genome Institute, Shanghai University, Shanghai 200444, China.

[c] University of Technology Sydney, School of Mathematical and Physical Sciences, Ultimo, New South Wales 2007, Australia.

[d] Department of Materials Science, University of Milan-Bicocca, Via Roberto Cozzi 55, 20125 Milan, Italy.

**Synopsis** The planar hexagonal phase of ZnO is significant in understanding ferroelectricity in this materials class, and for ZnO is well established in layered systems. In 2010 it was reported as freestanding nanocrystals, but the fitted lattice parameters differed from predictions by up to 16 %; we resolve this anomaly to provide fundamental understanding of ferroelectric switching.

**Abstract** The planar hexagonal phase of ZnO, known as h-ZnO, g-ZnO, α-ZnO, the $B_k$ structure, the 5-5 phase, the α-BN phase, etc., has $P6_3/mmc$ symmetry and is implicated in ferroelectric switching mechanisms for wurtzite-ZnO. It is well-known in thin films on substrates and to be stabilized by external pressure, but critical is its possible existence in high-purity nanocrystals under ambient conditions. Indeed, a crystal structure has been reported, but this work remains controversial as first-principles calculations predict very different structural properties. Herein, the original experimental data is re-refined, through phase-shift determination and Morlet wavelet transformation, that molecular dynamics simulations associate with a $P6_3/mmc$ structure with lattice parameters at room temperature of $a$ = 3.45±0.02 Å and $c$ = 4.46±0.02 Å. These values are 0.35 Å and 0.80 Å, respectively, larger than those previously reported and in good agreement with computational predictions. This confirms that ZnO nanocrystals can form a metastable planar hexagonal phase. It provides key information pertaining to polarization switching in ZnO, its derivatives, and general wurtzite-structured materials.

### 1. Introduction

Lizandra Pueyo et al. (Lizandara Pueyo *et al.*, 2010) have reported nanocrystals of ZnO with purity in excess of 99% that display $P6_3/mmc$ symmetry (Fig. 1a). They showed them to be metastable, converting to the wurtzite ($P6_3mc$) phase w-ZnO (Fig. 1b) at temperatures in excess of 200 °C. Qualitative evidence presented to support the $P6_3/mmc$ symmetry included: absorption spectroscopy, Raman spectroscopy, powder X-Ray diffraction (PXRD), X-ray absorption near-edge spectra (XANES), extended X-ray absorption fine structure (EXAFS), and high-resolution transmission electron microscopy (HRTEM). They also performed quantitative structural analysis of the EXAFS data to determine the lattice constants and interatomic distances. These results were then shown to be consistent with the PXRD data.

In detail, their analysis (Lizandara Pueyo *et al.*, 2010) revealed the structure to display trigonal bipyramidal coordination for both the Zn and O atoms (Fig. 1a), with in-plane Zn-O bond distances of 1.791 Å and out-of-plane distances of 1.928 Å, leading to hexagonal unit-cell vectors $a = b = 3.099$ Å and $c = 3.858$ Å (Table 1). Because of the trigonal bipyramidal coordination, this phase is often called the "5/5" or "5-5" phase (Lizandara Pueyo *et al.*, 2010, Zagorac *et al.*, 2012). Notably, the reported inter-planar spacing is extremely contracted compared to the value of 5.2057 Å for w-ZnO (Fig. 1b) (Schreyer *et al.*, 2014). If the $P6_3/mmc$ phase had the same interlayer spacing as found in w-ZnO, then the ZnO planes would be separated from each other by van der Waals bonding distances, and the coordination of the Zn and O atoms would be regarded as being trigonal planar, akin to the structure of hexagonal boron nitride (h-BN, Fig. 1c). It is therefore also common to label the $P6_3/mmc$ phase as either "α-BN", "h-ZnO", or "α-MgO" (Zagorac *et al.*, 2012), as well as "graphitic-ZnO" (Kulkarni *et al.*, 2005), "HX" (Kulkarni *et al.*, 2006), "gZnO" (Yadav *et al.*, 2021), "α-ZnO" (Wei *et al.*, 2011), and BN-ZnO (Zhang & Schleife, 2018), as well as the "$B_k$" structure (Molepo & Joubert, 2011). Accurate structural measurements, capable of differentiating between the structures portrayed in Fig. 1a and Fig. 1c, are therefore critical to the understanding of the properties of h-ZnO.

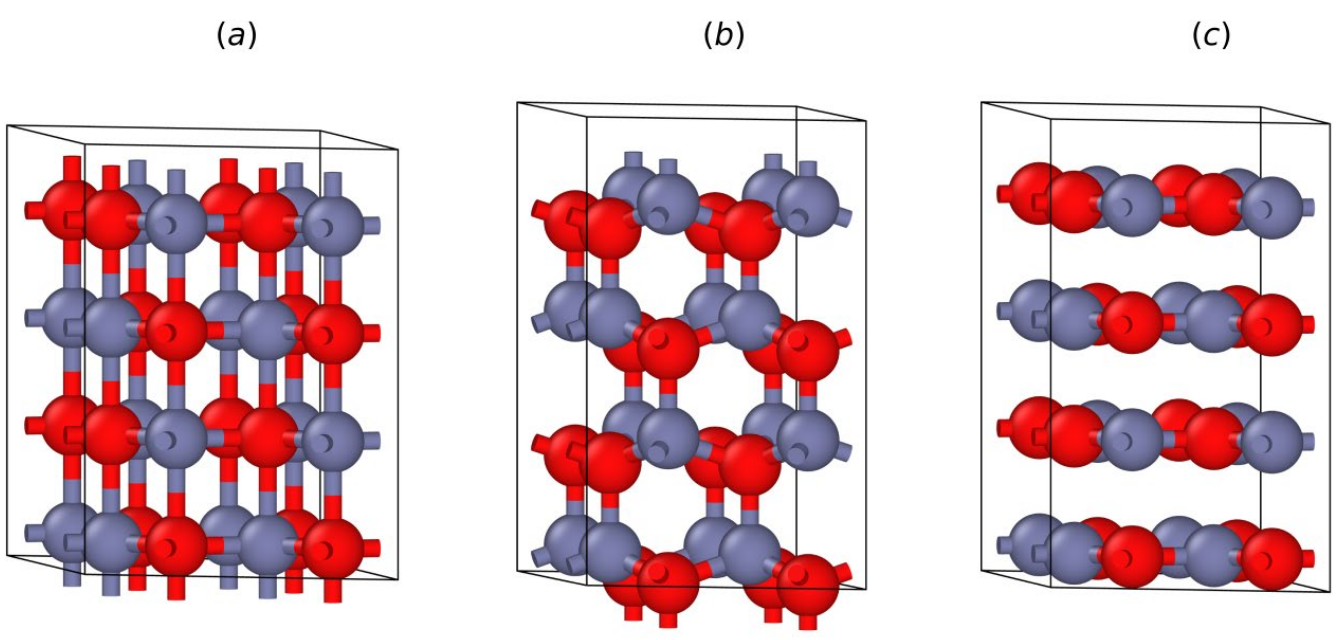

**Figure 1** Structures of: (a) h-ZnO (low volume observed trigonal bipyramidal structure), (b) w-ZnO (high volume observed tetrahedral structure), and (c) h-ZnO depicted at the observed lattice vectors of w-ZnO (high volume hypothetical h-BN structure).

**Table 1** Observed and first-principles calculated lattice parameters for h-ZnO at $T = 0$ K.

| Method | Reference | $a = b$ (Å) | $c$ (Å) |
|---|---|---|---|
| Original obs. | (Lizandara Pueyo *et al.*, 2010) text | 3.099 | 3.858 |
| Original obs. | (Lizandara Pueyo *et al.*, 2010) SI | 3.31 | 4.12 |
| LDA | (Rakshit & Mahadevan, 2011) | 3.371 | 4.459 |
| PBE | (Rakshit & Mahadevan, 2011) | 3.45 | 4.62 |
| HSE06 | (Kim, 2012) | 3.425 | 4.512 |
| HF | (Zagorac *et al.*, 2014) | 3.48 | 4.46 |
| B3LYP | (Zagorac *et al.*, 2014) | 3.48 | 4.54 |
| Revised obs. | This work | 3.45±0.02 | 4.46±0.02 |

Of significance, w-ZnO (Fig. 1b) is a useful high-bandgap high-polarization material but is not regarded as being a ferroelectric as, to date, no process has been found that can interconvert its polarization states. In the proposed concerted mechanism for ferroelectric switching of w-ZnO, the O atoms translate vertically from one ferroelectric form (Fig. 1b) to form h-ZnO (Fig. 1a), on route to the other ferroelectric form. Typically, h-ZnO is considered as the high-energy transition state that controls this process, and hence the claim (Lizandara Pueyo *et al.*, 2010) that freestanding h-ZnO nanocrystals can exist in a metastable phase challenges current understanding. As doping of ZnO by Mg has recently been shown to facilitate polarization switching (Ferri *et al.*, 2021) through a related concerted mechanism (Huang *et al.*, 2022), the nature of h-ZnO becomes significant to the basic understanding of ferroelectric switching in these, and indeed all, wurtzite-structured materials (Zagorac *et al.*, 2012, Zagorac *et al.*, 2013, Adhikari & Fu, 2019, Huang *et al.*, 2022, Ferri *et al.*, 2021).

To date, there has not been general acceptance of h-ZnO as a metastable intermediate in the concerted ferroelectric switching of w-ZnO. Firstly, calculations have predicted h-ZnO in infinite crystals to form a transition state along the polarization-switching pathway (Kim, 2012), with stabilization of h-ZnO predicted to occur only at high applied pressures (Molepo & Joubert, 2011, Su *et al.*, 2015, Wang *et al.*, 2015, Nakamura *et al.*, 2016, Zhang & Schleife, 2018, Wang *et al.*, 2022, Adnan *et al.*, 2025).

Although such calculations may not be reliable, more significant issues arise concerning the reported structure (Lizandara Pueyo *et al.*, 2010) of h-ZnO. From the qualitative perspective, the reported Zn-O separation of 1.791 Å is extremely short as Zn-O bond lengths usually exceed 1.9 Å. Even though ZnO is structurally flexible and can "tolerate huge distance variations" (Fischer *et al.*, 2023), this result does not appear to be reasonable. From the quantitative perspective, first-principles calculations, using density functional theory (DFT) (Rakshit & Mahadevan, 2011, Kim, 2012, Rakshit & Mahadevan, 2012, Molepo & Joubert, 2011, Su *et al.*, 2015, Zhang & Schleife, 2018, Zagorac *et al.*, 2014) or Hartree-Fock (HF) theory (Zagorac *et al.*, 2014), do not support such a bond-length contraction. They concurrently predict much larger lattice vectors of $a \sim 3.4 - 3.5$ Å instead of 3.099 Å, and $c \sim 4.4 - 4.6$ Å instead of 3.858 Å, see Table 1. As a result of these controversies, the very existence of a metastable h-ZnO phase of high purity ZnO under ambient conditions remains in doubt.

Layered structures containing h-ZnO, mostly supported on surfaces, are well established experimentally (Tusche *et al.*, 2007, Yadav *et al.*, 2021), with their existence supported by DFT calculations (Claeyssens *et al.*, 2005, Tu & Hu, 2006, Zhang & Huang, 2007, Das *et al.*, 2014, Freeman *et al.*, 2006). Such structures are important in their own right, with applications including hydrogen storage (Si *et al.*, 2011) and thermovoltaics (Li *et al.*, 2013), but are only peripherally relevant to bulk w-ZnO polarization switching. Concerning polarization switching, of note, the interlayer spacing in the bulk limit was predicted (Claeyssens *et al.*, 2005) by PW91 (Claeyssens *et al.*, 2005) calculations to be 4.10 Å, with initially observed double-inter-layer spacings being of order 4.2 Å – 4.8 Å (Tusche *et al.*, 2007) and modern measurements indicating 4.20 Å (Yadav *et al.*, 2021). Molecular dynamics simulations using empirical force fields also support the formation of h-ZnO in nanostructures (Kulkarni *et al.*, 2005, Kulkarni *et al.*, 2006).

Despite the large structural differences between different observations and most predictions, strong support for the experimental identification of nanocrystalline h-ZnO comes from comparison of observed and calculated spectroscopic properties. The experiments of Lizandra Pueyo et al. (Lizandara Pueyo *et al.*, 2010) match both *GW*/Bethe-Salpeter (Zhang & Schleife, 2018, Kang *et al.*, 2019) and time-dependent DFT (TDDFT) (Kang *et al.*, 2019) electronic spectral simulations, as well as Raman spectral simulations (Su *et al.*, 2015).

Lizandra Pueyo et al. (Lizandara Pueyo *et al.*, 2010) originally reported challenges to the interpretation of the critical EXAFS data used in their quantitative analysis. They reported in their supporting information two possible interpretations, both of which are indicated in Table 1 and can be seen to be significantly different. We pursue this feature, obtaining an unambiguous structure through by accurately determining the phase shift, Morlet wavelet transformation (WT), and molecular dynamics simulations of thermal structural effects.

## 2. Methods

First, the EXAFS spectrum is extrapolated using a pseudo-Voigt function (Limandri et al., 2008). Then the phase shift is determined using the method of Lee et al. (Lee & Beni, 1977), and the pair distribution function (PDF) determined. Morlet WT is then used to establish correlations between peaks in the real and reciprocal spaces (Timoshenko & Kuzmin, 2009) to provide an authoritative peak assignment. The obtained structure is then interpreted using molecular dynamics simulations at T = 293 K and fixed volume, using the GRACE-FS-OMAT force field (Bochkarev et al., 2024) using LAMMPS (Thompson et al., 2022), with the anticipated PDF peaks simulated using the Larch package (Newville, 2013). The MD time step used was 1 fs, initial trajectories at constant energy and volume were ran for 10 ps, then final trajectories ran for 100 ps. This allows the structure to be estimated at 0 K for easy comparison with most computational predictions.

## 3. Results

The original EXAFS spectrum (Lizandara Pueyo *et al.*, 2010) was digitized and is shown in Fig. 2, wherein a pseudo-Voigt function (Limandri *et al.*, 2008) is used to extrapolate the data into the unobserved part of the spectrum below $k$ = 3 Å$^{-1}$, as well as to reduce noise at high $k$. Critical to the data analysis is the determination of the phase shift that is induced by X-ray absorption (Lee & Beni, 1977, Lee *et al.*, 1981), a process that is traditionally performed empirically. Instead, following Lee et al. (Lee & Beni, 1977), the extrapolated EXAFS data is forward Fourier transformed, weighted by a windowing function, and then backward Fourier transformed (Fig. 2). In Supporting Information (SI) Figs. S1 and S2, various possible windowing functions are considered, and the Kaiser-Bessel function was selected for this purpose.

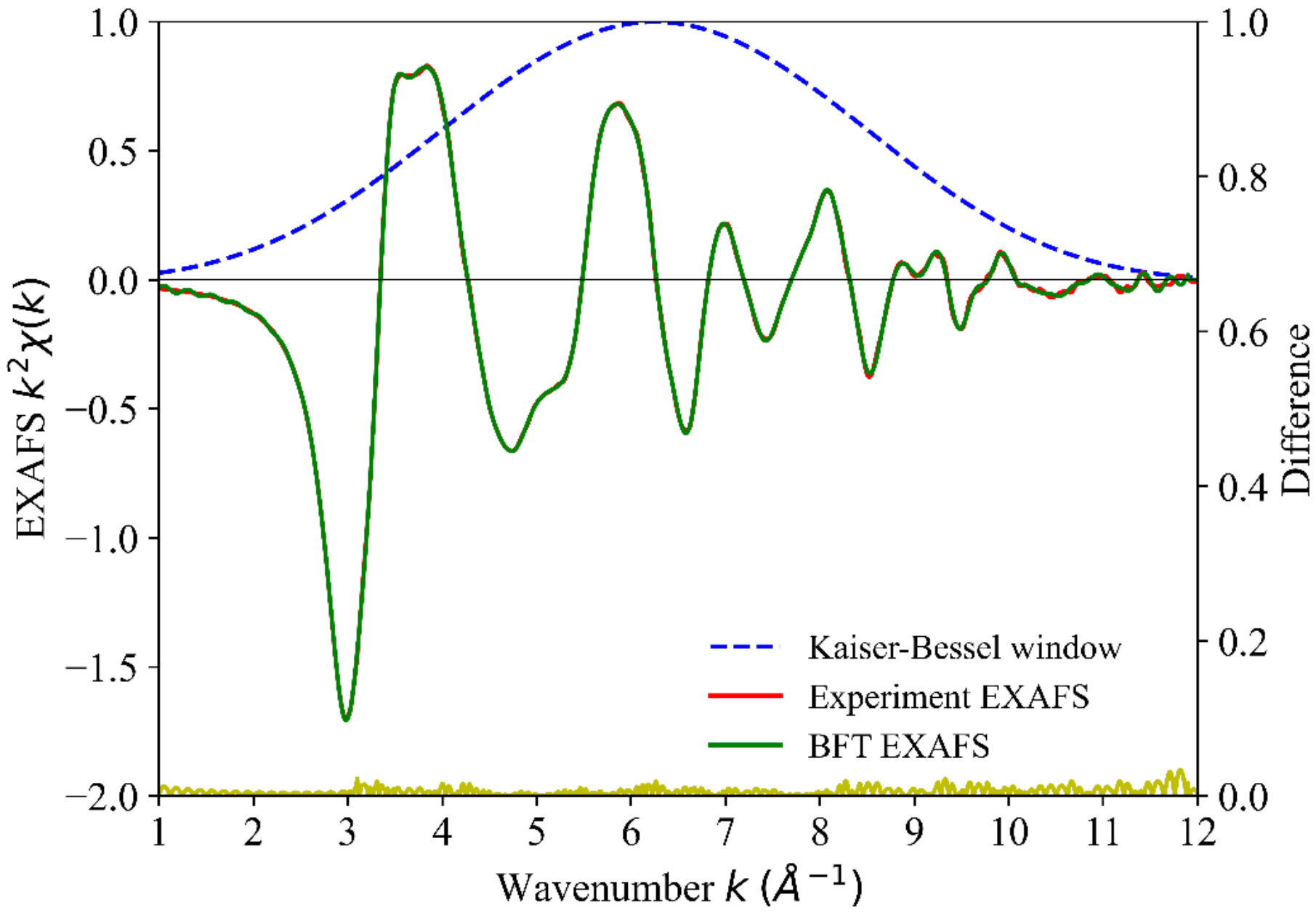

**Figure 2** The observed *k*-weighted EXAFS spectrum of h-ZnO (Lizandara Pueyo *et al.*, 2010) (red) is extrapolated using a pseudo-Voight function, forward Fourier transformed, weighted by a Kaiser-Bessel window (blue) and then backward Fourier transformed (BFT) to produce an expanded, noise-reduced, spectrum (green dashed). The differences, representing the noise reduced by the Fourier-transformation procedure, are shown in yellow using an expanded *y*-scale.

This Fourier-transformation process both reduces noise and allows the phase shift to be determined from the phase function $\phi(k)$, where

$$\phi(k) = \tan^{-1}[\mathrm{Im}(k)/\mathrm{Re}(k)] \qquad (1)$$

and the imaginary and real parts of EXAFS are calculated from the backward Fourier transform. The impact of the phase shift is to translate the perceived peaks in the PDF are by (Lee & Beni, 1977, Lee *et al.*, 1981)

$$\Delta R = \frac{-1}{2}\frac{\mathrm{d}\phi}{\mathrm{d}k} \qquad (2)$$

In SI Fig. S3, the phase function is plotted and shown to have a linear variation with $k$, allowing a positional shift of $\Delta R$ = 0.613 Å to be determined from Eqn. (2). This shift should be insensitive to phase and environment, and is consistent with values observed for w-ZnO variants of 0.6–0.8 Å (Neamtu *et al.*, 2010).

The final PDF resulting from this process (Lee & Beni, 1977, Lee *et al.*, 1981) is shown in Fig. 3. Its peaks in real space indicate inter-atomic distances, with key values labelled $r_1$ - $r_4$, but these features are also contaminated with backscattering contributions arising from interatomic interactions from within different coordination shells. To identify the primary origins of these features and establish an authoritative peak assignment, Morlet WT analysis is performed on the Zn K-edge EXAFS data (Timoshenko & Kuzmin, 2009). During this procedure, the wavelet parameters η = 8 and σ = 1 were used to provide a reasonable balance of resolution between $k$-space and real space (Timoshenko & Kuzmin, 2009). The results are shown in Fig. 4, which provides a correlation between the peaks in reciprocal space (Fig. 2) with those in real space (Fig. 3).

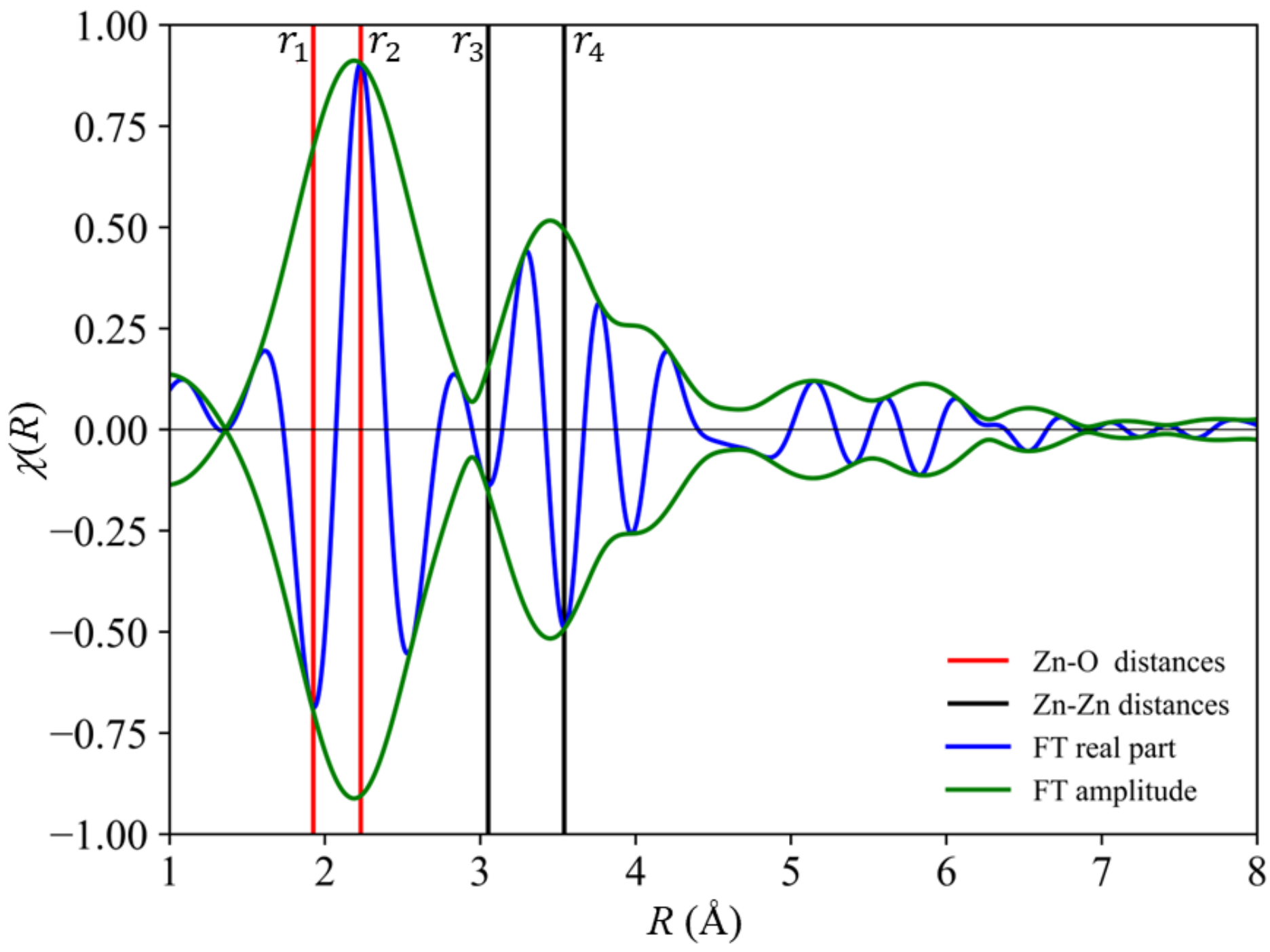

**Figure 3** The real (blue) and the total (green) PDF components obtained from the EXAFS data for h-ZnO, after correction for the phase shift. The red and black lines indicating key interatomic distances $r_1$- $r_4$.

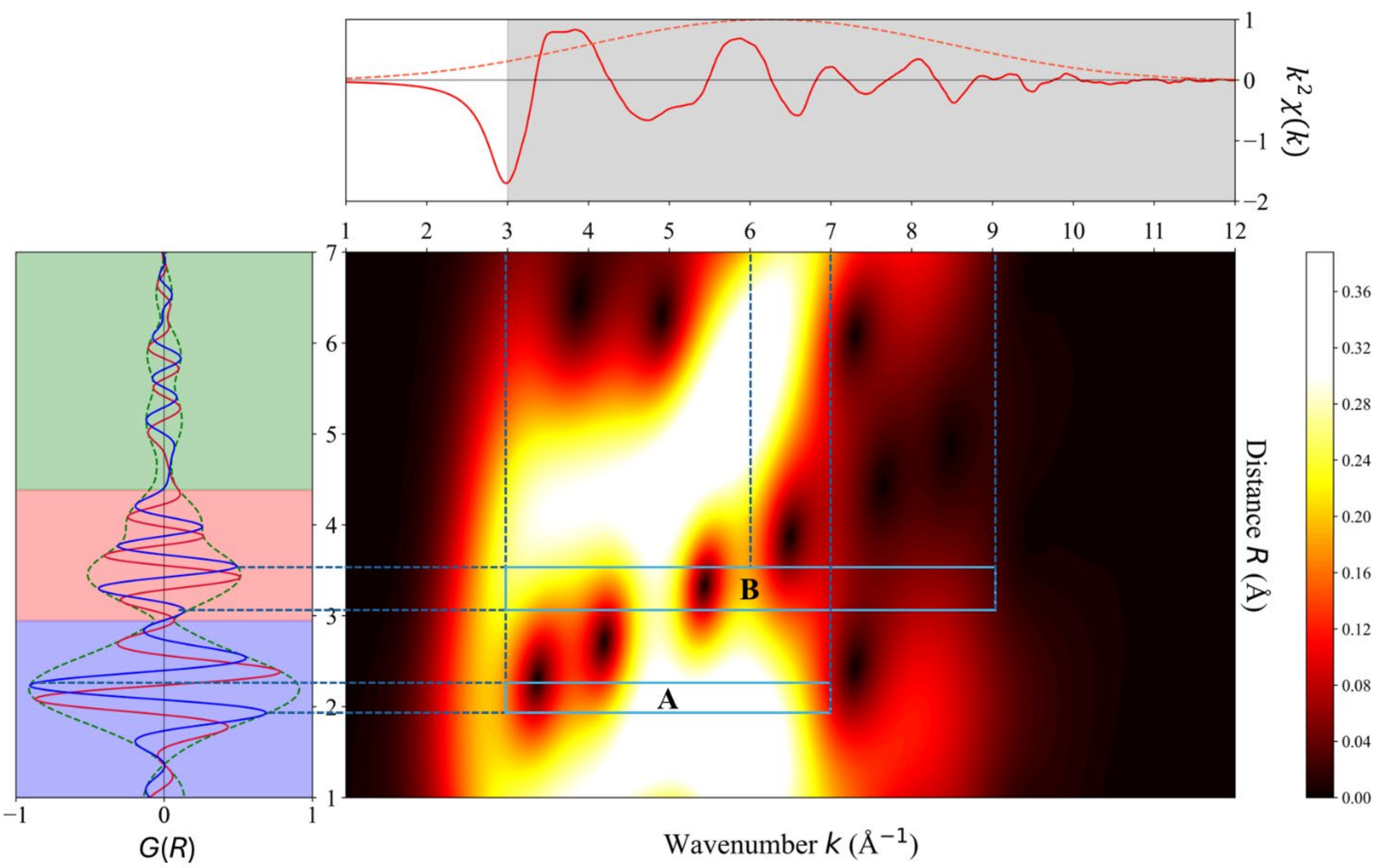

**Figure 4** The central image shows the wavelet-transformed EXAFS spectrum of h-ZnO, which correlates the EXAFS signal $k^2\chi(k)$ in reciprocal space $k$ (top, Fig. 2) with that as Fourier-transformed into real space $G(R)$ (left, Fig. 3). Peaks are indicative of either the indicated interatomic distances or else multi-scattering paths (blackened regions of signal loss), with key regions labelled A and B.

The WT contour plot in Fig. 4 shows a concentrated intensity distribution in the region labelled "A" embodying $R$ ~ 1.9–2.3 Å and $k$ ~ 3 - 7 Å$^{-1}$. In this region, the contour plot mostly exhibits a smooth elliptical distribution, without significant modulation, that is consistent with the simple two-body scattering path characteristics of oxygen as a light backscattering atom ($Z$ = 8). Region A is therefore associated with Zn-O nearest-neighbor scattering and hence the peaks at distances $r_1$ and $r_2$ in Fig. 3 correspond to Zn-O distances. From the observed intensities, the first-shell coordination number is estimated to be 5.03 ± 0.1 (Fig. S2), as expected for the trigonal-bipyramidal structure of h-ZnO.

In contrast, in the region labelled "B" in Fig. 4 embodying $R \sim 3.1$–$3.5$ Å and $k = 3$–$9$ Å$^{-1}$ there is observed pronounced intensity modulation in $k$-space, with a primary region spanning $k \sim 3$-$6$ Å$^{-1}$ with a hole in the middle, and a secondary peak at $k \sim 9$ Å$^{-1}$. This is consistent with the characteristics of Zn as a heavy backscattering atom ($Z$ = 30) and indicates that this region contains superimposed contributions from Zn-Zn single scattering and Zn-O-Zn three-atom multiple scattering paths. Hence the peaks at distances $r_3$ and $r_4$ in Fig. 3 correspond to Zn-Zn separations.

This interpretation leads to the identification of the peaks highlighted in Fig. 3 as: that at $r_1$ = 1.915 Å corresponds to the in-plane Zn-O distance, that at $r_2$= 2.23 Å corresponds to the inter-layer Zn-O distance (half of the lattice vector $c$), that at $r_3$ = 3.05 Å corresponds to the inter-layer Zn-Zn distance, and that at $r_4$ = 3.45 Å corresponds to the in-plane Zn-Zn distance (and hence the lattice vector $a$). These results are broadly consistent with expectations based on the computationally optimized structures listed in Table 1 and differ significantly from the options proposed originally (Lizandara Pueyo *et al.*, 2010).

Four unique interatomic distances are thus determined, whereas only two features, the lattice vector lengths $a$=$b$ and $c$ determine structures of $P6_3/mmc$ symmetry. Therefore two relationships must be obeyed in order to confirm this structure, as detailed in Table 2. The expected relationships are found to be in error by up to 7%, a considerable value that challenges the symmetry assignment. To understand this, molecular dynamics simulations were performed at $T$ = 293 K and constant volume using the GRACE-FS-OMAT force field (Bochkarev *et al.*, 2024) using LAMMPS (Thompson *et al.*, 2022). Using lattice parameters of $a$ = 3.45 Å and $c$ = 4.46 Å, these simulations, without inclusion of multiple-scattering events, the real component of the PDF $\chi(R)$ was calculated yielded (Newville, 2013) and is shown in SI Fig. S3 where it is shown to be in good agreement with the observed data. From this, values for $r_1$ - $r_4$ were extracted and listed in Table 2. These results are in good agreement with those observed, indicating that thermal motion is responsible for the apparent observed lowering of symmetry from $P6_3/mmc$. The structure is therefore confirmed to be h-ZnO.

**Table 2** Expected ratios of interatomic bond distances for *P6₃/mmc* symmetry, compared with those obtained from the EXAFS data and those obtained from MD simulations at $T$ = 298 K using lattice vectors $a$ = 3.45 Å and $c$ = 4.46 Å.

| ratio | expected | observed | MD |
|---|---|---|---|
| $r_4/r_1$ | $\sqrt{3} = 1.73$ | 1.80 | 1.79 |
| $(r_1^2 + r_2^2)/r_3^2$ | 1 | 0.93 | 0.92 |

To robustly estimate systematic uncertainties, we varied the shape parameter $\beta$ in the Kaiser-Bessel windowing function from 3 to 15 (a range chosen to fully encompass the mathematical trade-off between spatial resolution and spectral leakage), and tested six alternative windowing functions (Rectangular, Hamming, Norton-Beer, Gaussian, Right tail, and Left tail), see SI Figs. S1-S2. The standard deviation of the structural parameters derived from these comprehensive variations, combined with uncertainties in the effects of thermal motion, yields estimated uncertainties for the structural analysis. Uncertainties with respect to the determination of the phase shift and reading the peaks in Fig. 3 are small in comparison to the effect of these method variations. This analysis results in the room-temperature lattice parameters for h-ZnO of a = 3.45 ± 0.02 Å and c = 4.46 ± 0.02 Å (Table 1). These results are in quantitative agreement with the computed structures.

A cif file describing the re-refined structure is provided in SI. These results confirm that, for the synthesized nanocrystals (Lizandara Pueyo *et al.*, 2010), h-ZnO presents as a metastable phase and hence is not a transition state along the pathway for concerted ferroelectric switching of w-ZnO.

#### 4. Conclusions

The long-standing controversy concerning the identification and properties of the h-ZnO nanocrystalline phase has been resolved. Under suitable conditions, a metastable phase can be isolated that has properties similar to those observed for nanolayered ZnO structures stabilized by substrate surfaces.

Heating of the observed nanocrystals led to formation of a wurtzite structure (Lizandara Pueyo *et al.*, 2010). If an electric field of sufficient magnitude could be applied to these crystals, then the crystals would surmount a reaction barrier and convert back to h-ZnO; this would remain metastable after the applied field was removed. Further increasing of the field strength could then lead to ferroelectric switching, as has been observed for say $Zn_{0.5}Mg_{0.5}O$ (Huang *et al.*, 2022).

It remains to be established if this result applies only to certain nanocrystals or else is expected to be general for all ZnO crystals. Previous computational approaches have indicated that the result affirmed here for the nanocrystals studied does not apply to infinite crystals. In future work (Zhang *et al.*, 2026), these computational results need to be tested for robustness. In addition, calculations need to be applied to model the synthesized nanocrystals (Lizandara Pueyo *et al.*, 2010) in their external environment. Such studies are relevant to the understanding of concerted polarization switching in all wurtzite-structured materials.

**Acknowledgements** We thank the National Natural Science Foundation of China (12404276, 12347164), the China Postdoctoral Science Foundation (2024T170541, GZC20231535), the China Scholarship Council (No.202406890093), and the Australian Research Council Centre of Excellence in Quantum Biotechnology (CE230100021).

**Conflicts of interest** There are no conflicts of interest.

**Data availability** The optimized cif file is provided in Supporting Information.

# Supporting information

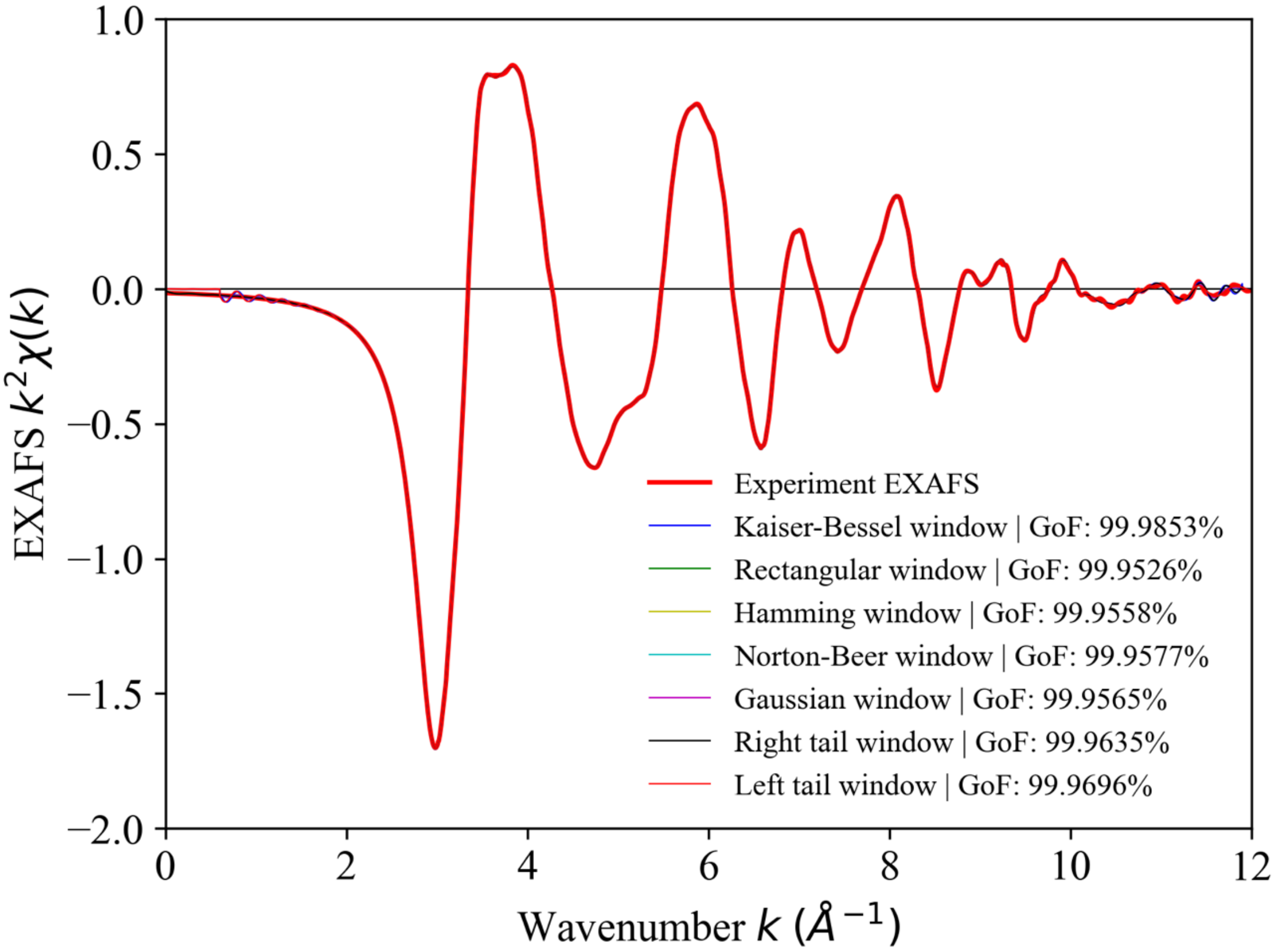

**Figure S1** EXAFS spectrum showing the $k^2$-weighted $\chi(k)$ function plotted against wavenumber $k$ ($Å^{-1}$) in the range of 0-12 $Å^{-1}$. The red curve represents experimental EXAFS data, and the colored curves represent calculated results that are all barely distinguishable from it. The goodness of fit (GoF) of backward FT EXAFSs using seven different window functions for Fourier transformation are listed: Kaiser-Bessel window (GoF: 99.9853%), Rectangular window (GoF: 99.9526%), Hamming window (GoF: 99.9558%), Norton-Beer window (GoF: 99.9577%), Gaussian window (GoF: 99.9565%), Right tail window (GoF: 99.9635%), and Left tail window (GoF: 99.9696%).

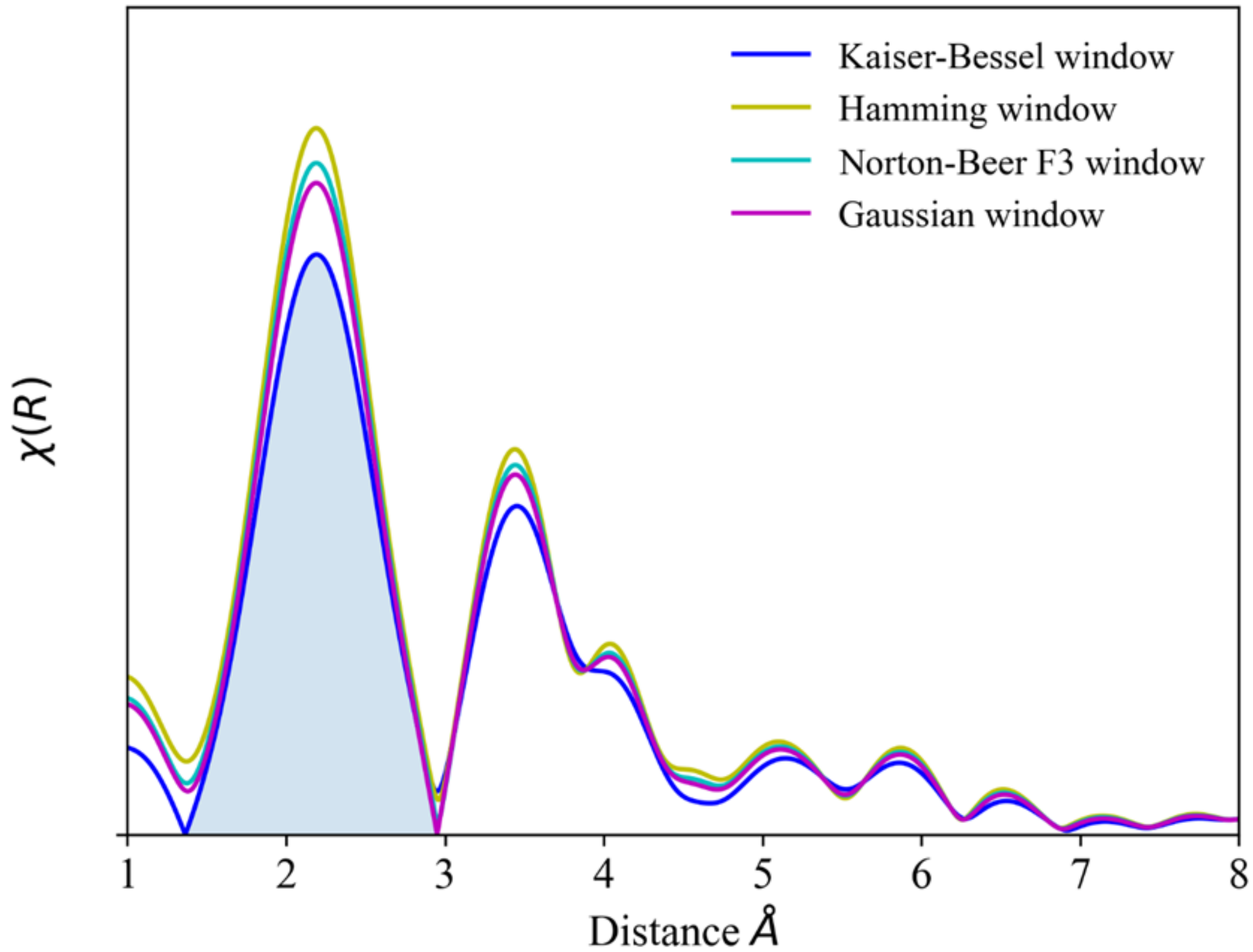

**Figure S2** Pair distribution functions derived from EXAFS data using four different windowing functions: Kaiser-Bessel, Hamming, Norton-Beer F3, and Gaussian. Integration of the first peak area (shaded) shows that the Kaiser-Bessel window produces a coordination number of 5.03 ± 0.1. This is close to the theoretical value of 5 for h-ZnO, demonstrating optimal balance between spectral leakage suppression and signal fidelity.

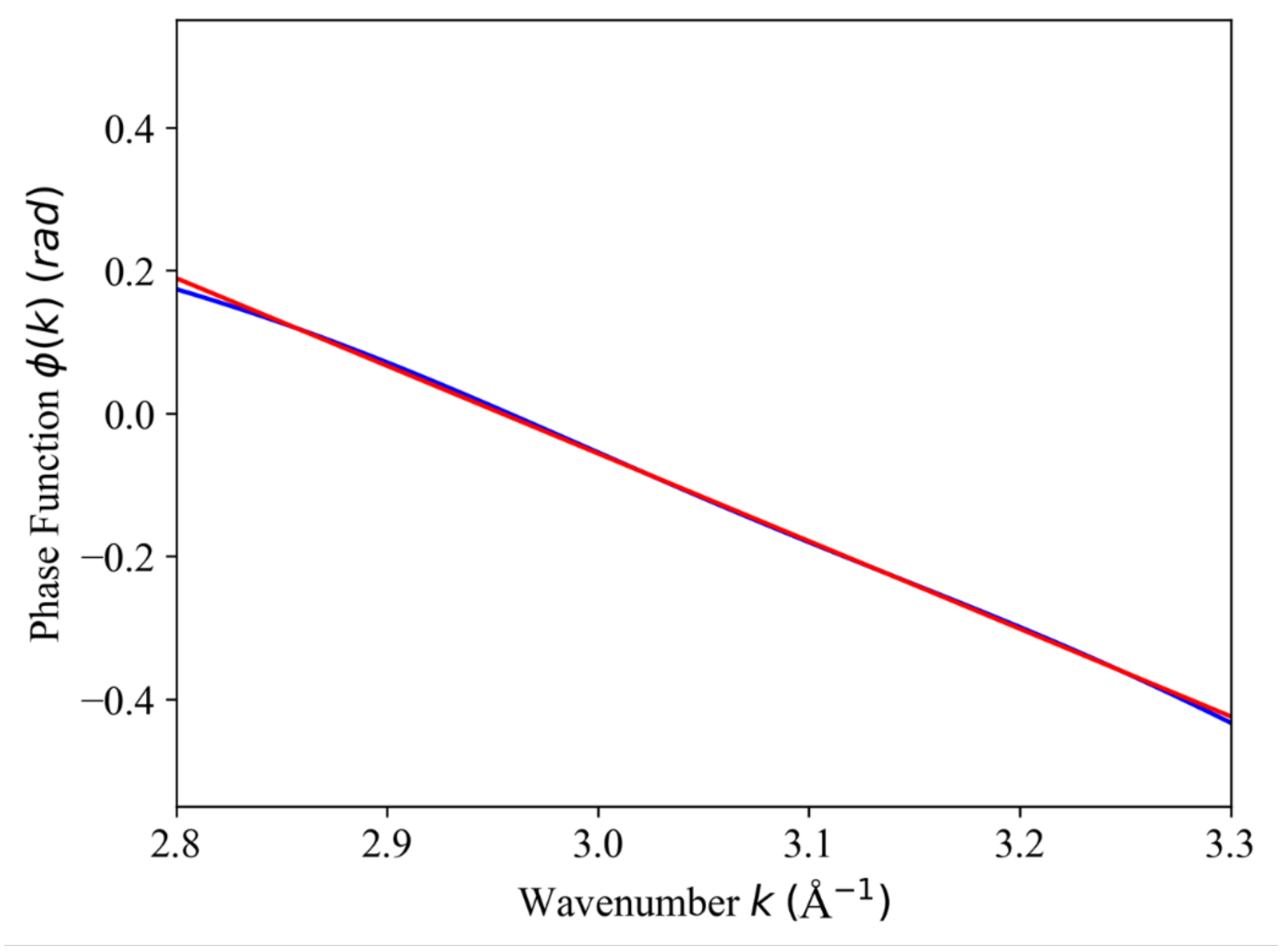

**Figure S3** Figure S3 The phase function (Eqn. (1)) (blue) is plotted and fitted to a straight line (red) to extract the derivative ${}^{\mathrm{d}\phi}/_{\mathrm{d}k}$ for use in Eqn. (2).

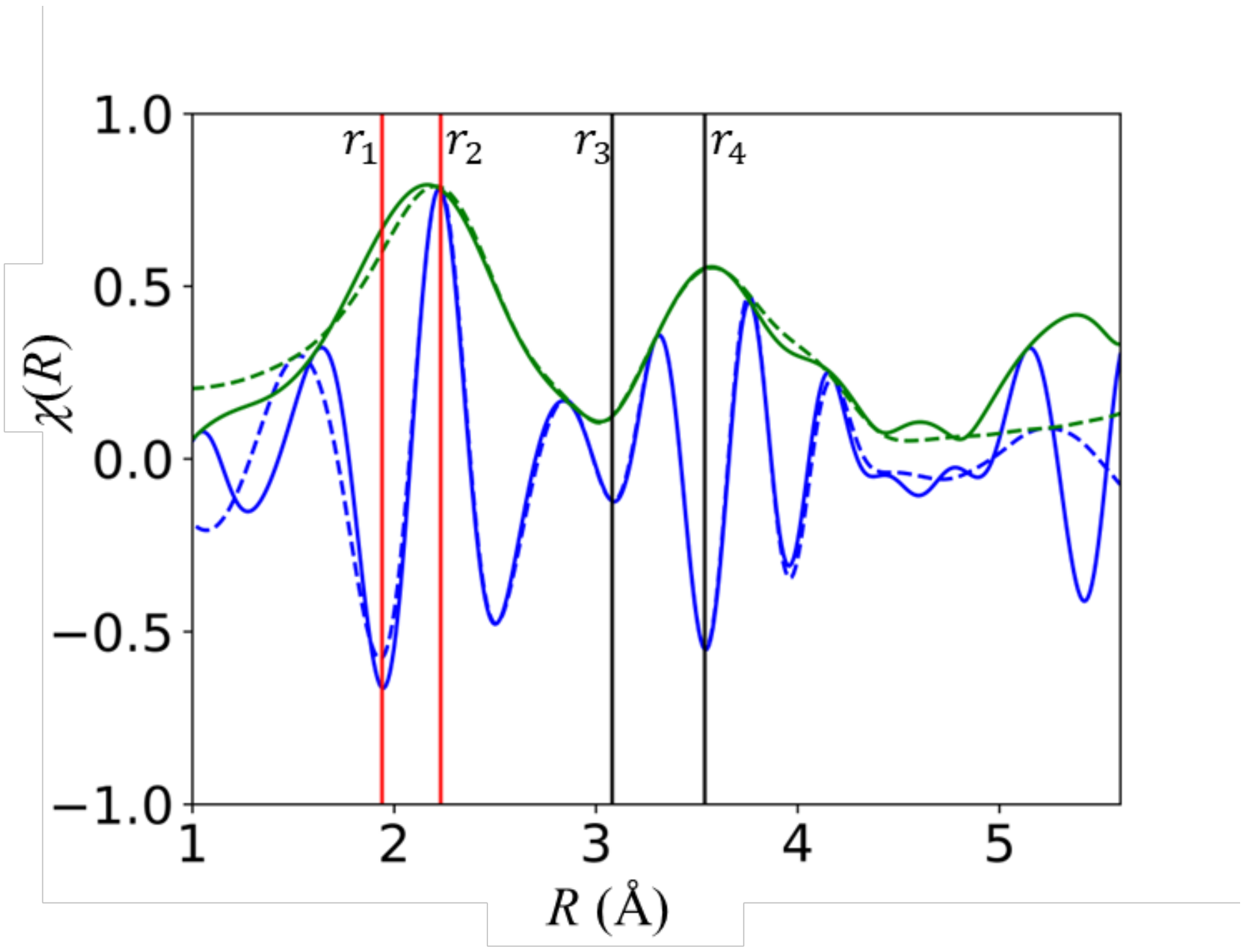

**Figure S4** The PDF $\chi(R)$ obtained from the MD simulations performed at constant temperature and volume is used to determine thermal interatomic distances $r_1$ - $r_4$. This was obtained by Fiuriel-transformation of the EXAFS spectrum simulated from the simulation results using the method of Larch (Newville, 2013). Green- total PDF, blue- real part only; solid lines- MD simulation, dashed lines- observed from Fig. 3.